\documentclass[preprint,12pt]{elsarticle}

\usepackage{amssymb}
\usepackage{xcolor} % For color customization
\usepackage{titlesec} % For customizing section formatting

\definecolor{mytextcolor}{gray}{0.4}

\newcommand{\subsubsubsection}[1]{%
    \par\refstepcounter{paragraph}%
    \addcontentsline{toc}{paragraph}{\theparagraph\quad #1}% Add to TOC (optional)
    \noindent\textcolor{mytextcolor}{\normalfont\normalsize\textbf{\theparagraph\quad #1}}\par\nobreak%
    \vspace{1.0ex plus .2ex}%
}

\usepackage{booktabs}
\usepackage{array}
\usepackage{comment}
\usepackage{amsmath}
\DeclareMathOperator*{\argmin}{arg\,min}

\usepackage{caption}

\usepackage{graphicx}
\usepackage{caption}
\usepackage{subcaption}
\usepackage{kantlipsum} %dummy text 
\usepackage{float}

\UseRawInputEncoding

\usepackage{tabularray}

\usepackage{amsmath}
\makeatletter
\let\reftagform@=\tagform@
\def\tagform@#1{\maketag@@@{(\ignorespaces\textcolor{blue}{#1}\unskip\@@italiccorr)}}
\renewcommand{\eqref}[1]{\textup{\reftagform@{\ref{#1}}}}
\makeatother
\usepackage{hyperref}
\hypersetup{colorlinks=true}
\newcommand{\norm}[1]{\left\lVert #1 \right\rVert}

\usepackage{lineno}
\usepackage{subcaption}
\newcommand{\minus}{\scalebox{0.75}[1.0]{$-$}}

\usepackage{graphicx}
\usepackage{caption}
\usepackage[skip=0.5ex]{subcaption}
\usepackage[parfill]{parskip}

\begin{document}

\begin{frontmatter}



\title{Training Compute-Optimal Vision Transformers for Brain Encoding}

\author[label1,label2,label3]{Sana Ahmadi\footnote{Corresponding author, sana.ahmadi@mail.concordia.ca}}
\author[label2,label3]{Fran\c{c}ois Paugam\footnote{francois.paugam@umontreal.ca}}
\author[label1]{{Tristan Glatard}\footnote{tristan.glatard@concordia.ca}}
\author[label2,label3]{{Pierre Lune Bellec}\footnote{pierre.bellec@criugm.qc.ca}}
\address[label1]{Department of Computer Science and Software Engineering, Concordia University, Montreal, QC, Canada}
\address[label2]{Université de Montréal, Montréal, QC, Canada}
\address[label3]{Centre de Recherche de L'Institut Universitaire de Gériatrie de Montréal, Montréal, Canada}



%\address{California, United States}

\begin{abstract}
The optimal training of a vision transformer for brain encoding depends on three factors: model size, data size, and computational resources. This study investigates these three pillars, focusing on the effects of data scaling, model scaling, and high-performance computing on brain encoding results. Using VideoGPT to extract efficient spatiotemporal features from videos and training a Ridge model to predict brain activity based on these features, we conducted benchmark experiments with varying data sizes (10k, 100k, 1M, 6M) and different model configurations of GPT-2, including hidden layer dimensions, number of layers, and number of attention heads. We also evaluated the effects of training models with 32-bit vs 16-bit floating point representations. Our results demonstrate that increasing the hidden layer dimensions significantly improves brain encoding performance, as evidenced by higher Pearson correlation coefficients across all subjects. In contrast, the number of attention heads does not have a significant effect on the encoding results. Additionally, increasing the number of layers shows some improvement in brain encoding correlations, but the trend is not as consistent as that observed with hidden layer dimensions. The data scaling results show that larger training datasets lead to improved brain encoding performance, with the highest Pearson correlation coefficients observed for the largest dataset size (6M). These findings highlight that the effects of data scaling are more significant compared to model scaling in enhancing brain encoding performance. Furthermore, we explored the impact of floating-point precision by comparing 32-bit and 16-bit representations. Training with 16-bit precision yielded the same brain encoding accuracy as 32-bit, while reducing training time by 1.17 times, demonstrating its efficiency for high-performance computing tasks.

\end{abstract}

\end{frontmatter}


%%

\section{Introduction}
\label{S:1}

Brain encoding aims to predict neural responses to stimuli by leveraging computational models. Traditionally, Convolutional Neural Networks (CNNs) such as AlexNet\cite{ImageNet}, VGG16 \cite{vgg16}, and ResNets \cite{Resnet} have been employed to mimic brain activity patterns, extracting semantic information from visual stimuli ~\cite{ vision2, vision3, H_DS, BTB, LSTM1, IM_1}. However, deeper CNN architectures have not consistently replicated brain-like responses across all regions and  recent advancements highlight the potential of transformers \cite{NLP_models, bertmodel, GPT_breain}. Transformer-based architecture \cite{Attention} offer several distinct advantages for brain encoding, including superior performance in capturing spatial and temporal features compared to CNNs, RNNs and LSTMs.  The attention mechanisms in transformers enhance the selective integration of visual inputs, which is crucial for understanding neural activity. Furthermore, generative self-supervised models demonstrate predictive capabilities comparable to supervised models. Finally, multi-modal architectures, such as VisualBERT \cite{Visualbert} and CLIP \cite{clip}, effectively leverage semantic correlations across different modalities, offering robust performance in visio-linguistic tasks. Inspired by these advancements, neuroscientists have developed transformer-based brain encoding models that significantly improve the accuracy of fMRI encoding across the entire brain \cite{Pereira, NLP_models, bertmodel, GPT_breain, Lanq-vision}.

In the pursuit of developing compute-optimal models, the interplay among computational resources, model dimensions, and dataset sizes holds paramount importance \cite{chinchilla,Scaling_law_vit, Scaling_law_Revisiting, Scaling_law_vision_transformers}. Scaling laws serve as fundamental frameworks elucidating the dynamics between a model's efficacy and these three pillars. For instance, the authors in \cite{chinchilla} delve into the optimal configuration of model size and training data for transformer language models based on computing resources budget. Their findings highlight that large-scale models often undergo undertraining due to an emphasis on scaling model sizes while keeping the training dataset size constant. By scaling both model size and training data size, the proposed model, Chinchilla, surpasses larger counterparts in performance metrics while utilizing fewer computational resources. Similarly, in \cite{Scaling_law_vit}, the authors advance scaling laws for vision transformers to propose compute-optimal model architectures regarding width and depth instead of solely focusing on the number of parameters.  

Recent research has extended scaling laws to brain encoding tasks, mirroring discussions found in studies of end-to-end language and vision models (trained from scratch on stimuli) alongside their brain encoding correlates. These studies aim to address fundamental questions such as: (1) How does the sample size to train transformers impact on brain encoding prediction accuracy? (2) How does brain encoding prediction accuracy vary with the parameter size (model size) of transformer? For example, a recent study \cite{Scaling_law_Alex} explored the effectiveness of larger language models, such as those from the OPT \cite{OPT} and LLaMA \cite{Llama} families (30B parameters), in predicting brain responses compared to traditional model GPT-2 (125M parameters). The results indicated a logarithmic relationship between brain prediction performance and model size. Similarly, scaling laws were observed with increasing training dataset, demonstrating that brain encoding prediction accuracy increases on a logarithmic scale with both the size of the training samples and the parameter size of the language models employed. The study in \cite{Scaling_law_vision_brain} explores the construction of a high-performance vision encoding model, assessing how changes in the sample size of the fMRI training set and the parameter size of vision models affect prediction accuracy. Various vision models with parameter sizes ranging from 86M to 4.3B were employed to extract features from stimuli presented to the subjects. Results demonstrate that increasing the training sample size and the parameter size of vision models enhances prediction accuracy of brain encoding according to the scaling law.

In this study, we investigate the optimization of vision transformers for brain encoding using the Shinobi dataset \cite{newshinobi}, which includes approximately 10 hours of fMRI data collected as subjects engaged with the Shinobi video game. This dataset provides a diverse range of cognitive engagements across different game levels, offering a rich foundation for brain encoding experiments. To extract efficient  spatiotemporal features from this dataset, we train an end-to-end VideoGPT~\cite{video_GPT} model with tens of millions of parameters. VideoGPT is able to capture complex spatial and temporal patterns in video data, which are essential for understanding brain responses to visual stimuli. The extracted features from VideoGPT are then paired with corresponding fMRI data, and we train a Ridge regression~\cite{Ridge1} model on these pairs to predict brain activity for unseen frames, enabling the decoding of neural responses to new visual stimuli. Our investigation is structured to address several key aspects: 1) exploring the effects of varying dataset sizes (10k, 100k, 1M, 6M) on brain encoding performance,  2) examining different model configurations such as hidden layer dimensions, number of layers, and attention heads, 3) assessing the impact of mixed precision on model training time and brain encoding accuracy.

\section{Materials and Methods}
\subsection{fMRI datasets}
 The fMRI Shinobi videogame dataset was collected in the context of the Courtois Neuromod Project. This game has been selected to effectively engage subjects with multiple cognitive components simultaneously, such as perception of the environment, strategic planning, decision making and taking action. In the Shinobi dataset, about 10 hours of fMRI data was recorded while the subjects play the Shinobi video game.  In each run, subjects played 3 levels in cycles and in the same order each time. These levels were: Level-1) corresponded to round 1 of the original game which included one mini-boss and one boss fight.  Level-4) corresponded to the beginning of round 4 of the original game which  included no mini-boss or boss fight. Level-5) corresponded to the beginning of round 5 of the original game, which included one mini-boss fight and no boss fight. Subject moved to the next level in two cases: they successfully completed a level, or lost three lives. The duration of each run is a minimum of ten minutes. A run was completed as soon as its duration exceeded 10 minutes and the participant completed a level, as was just defined. The duration of each run was thus variable, depending on the individual gameplay of the participant. As there are fixed order in the levels, Level-1 was repeated more frequently than Level-4 and Level-5.  For more information on this dataset, visit the CNeuroMod dataset documentation  page(\href{https://docs.cneuromod.ca/en/latest/DATASETS.html#shinobi} {CNeuroMod web page}).

\subsubsection{Participants}
   The Shinobi dataset includes fMRI time series collected on four participants in good general health, 2 women (sub-04, and sub-06) and 2 men (sub-01, sub-02).  All subjects also provided written informed consent to participate in this study, which was approved by the local research ethics review board (under project number CER VN 18-19-22) of the CIUSSS du Centre-Sud-de-l'Île-de-Montréal, Montréal, Canada.

\subsubsection{Magnetic resonance imaging}
Magnetic resonance imaging (MRI) was collected using a 3T Siemens Prisma Fit scanner and a 64-channel head/neck coil, located at the Unit for Functional Neuroimaging (UNF) of the Research Centre of the Montreal Geriatric Institute (CRIUGM), Montréal, Canada. Functional MRI data were collected using an accelerated simultaneous multi-slice, gradient echo-planar imaging sequence \cite{Setsompop,Xu} developed at the University of Minnesota, as part of the Human Connectome (HCP) Project \cite{Glasser}. The fMRI sequence used the following parameters: slice acceleration factor = 4, TR = 1.49s, TE = 37 ms, flip angle = 52 degrees, 2 mm isotropic spatial resolution, 60 slices, acquisition matrix 96x96. The structural data was acquired using a T1-weighted MPRAGE 3D sagittal and the following parameters: duration 6:38 min, TR = 2.4 s, TE = 2.2 ms, flip angle = 8 deg, voxel size = 0.8 mm isotropic, R=2 acceleration.  
For more information on the sequences used or information on data acquisition (including fMRI setup), visit the \href{https://docs.cneuromod.ca/en/latest/MRI.html#sequences} {CNeuroMod technical documentation} page. 

\subsubsection{Preprocessing}
All fMRI data were preprocessed using the fMRIprep pipeline version 20.2.3 \cite{Esteban}. We applied a volume-based spatial normalization to standard space (\allowbreak{MNI152 NLin2009cAsym}). Furthermore, a denoising strategy was applied to regress out the following basic confounds: (1) a 24-degrees of freedom expansion of the motion parameters, (2) a basis of slow time drifts (slower than 0.01 Hz). This step was implemented with the Nilearn maskers (see below) and the \texttt{load\_confounds} tool\footnote{\url{https://github.com/simexp/load_confounds}} (option \texttt{Params24}). A spatial smoothing with a 8 mm field-width-at-half-maximum and a Gaussian kernel was also applied with Nilearn prior to time series extraction. For each fMRI run, time series were also normalized to zero mean and unit variance (over time, for each voxel independently).

\subsubsection{Brain parcellation} 
 The preprocessed BOLD time series were averaged across all voxels in each parcel of a parcellation atlas, using the NiftiLabelsMasker
masker from Nilearn. We used the Multiresolution Intrinsic Segmentation
Template (MIST) \cite{MIST}. MIST provides a hierarchical decomposition of functional brain networks in nine levels, and we used here the largest
available resolution ($MIST_{AtOM}$). For each subject, the validation data is the data from the session 004, the test data is from session 005 and the training is the rest of the sessions. Details of fMRI data representation using $MIST_{AtOM}$ parcellation is presented in Table \ref{Table:Data_size}. 

\begin{table}[t]
\scriptsize
\begin{center}
\caption{Shinobi fMRI data representation using $MIST_{AtOM}$ parcellation}
\renewcommand{\arraystretch}{1.2} 
    \begin{tabular}{|c|c|c|}
    \hline 
    fMRI data representing & parcel-wise train  &  parcel-wise test \\
   \hline \hline
   shape of sub-01 data         &  (15313, 1095)    &(1513, 1095) \\
    \hline
    shape of sub-02 data        &  (17996, 1095)     & (829, 1095) \\
   \hline
   shape of sub-04 data       &    (14046, 1095)   &  (1867, 1095)  \\   \hline
   shape of sub-06 data     &  (13966, 1095)       & (1479, 1095)  \\  \hline
    Size   &           $\sim$151M  &   $\sim$12M    \\  \hline
    \end{tabular}
    \label{Table:Data_size}
\end{center}
\end{table}

\subsection{Generating Video Data}

The generation of video data for the CNeuromod Shinobi gameplay recordings involved transforming \textit{keypress} logs into visual frames that represent the gameplay experience. This process is pivotal for creating a dataset that captures not only the player's actions but also the corresponding visual responses from the game environment. The method employed integrates game emulation with keypress playback to achieve a structured dataset suitable for subsequent analysis.

The process begins with the setup of the  \textit{emulator} using the \textit{retro} library, which facilitates the playback of Sega Genesis games. Specifically, the emulator is configured to run \textit{"ShinobiIIIReturnOfTheNinjaMaster-Genesis"} leveraging custom integrations to ensure compatibility with the CNeuromod dataset. This step is critical as it establishes the foundation for accurately replaying gameplay sessions based on recorded inputs. Once the emulator is set up, the keypress logs, stored in \textit{.bk2} files, are loaded into the system. Each log file contains a sequential record of player inputs, and the emulator is reset to its initial state, ensuring it can accurately reflect the gameplay as originally experienced. This resetting process allows for the emulator to start from the exact point in the game where the recorded session began, thereby preserving the integrity of the gameplay dynamics. 

As the emulator replays each log, it simulates the exact sequence of actions taken by the player, advancing frame by frame. The gameplay environment's visual output is captured at each step, resulting in a series of images that represent the game state over time. To ensure consistency, the captured frames are resized to a standard resolution of $64\times64$ (or $128\times128$ ) pixels, maintaining their visual quality while fitting within the requirements of the dataset.

 \subsection{Brain encoding}
\label{gpt-extract features}
In this work, an end-to-end VideoGPT model was trained on the Shinobi dataset, which consists of more than 6 million frames totaling 309GB and includes 10-hours video recording of Shinobi gameplay across 4 subjects. After training the VideoGPT model, we used the trained model to extract spatio-temporal features from stimuli for brain encoding.

We specifically used the  \text{``attn\_stack.attn\_nets.4.post\_fc\_dp"} layer to represent the activations of VideoGPT. This layer was selected because the model includes 8 blocks, and we chose the fourth block to extract mid-level features, compare to the high-level features in later layers or low-level features in earlier layers. Additionally, we conducted a benchmark analysis among the layers in the 4th block and found that the \text{11attn\_stack.attn\_nets.4.post\_fc\_dp"} layer had lower activation size compared to other layers and provided a higher correlation for brain encoding (see Figure \ref{Brain_Pearson_Correlation_layers} and Table \ref{Table:layers}
in \ref{Appendix_1}).

\begin{figure}[t]
\centering
\includegraphics[width=12cm]{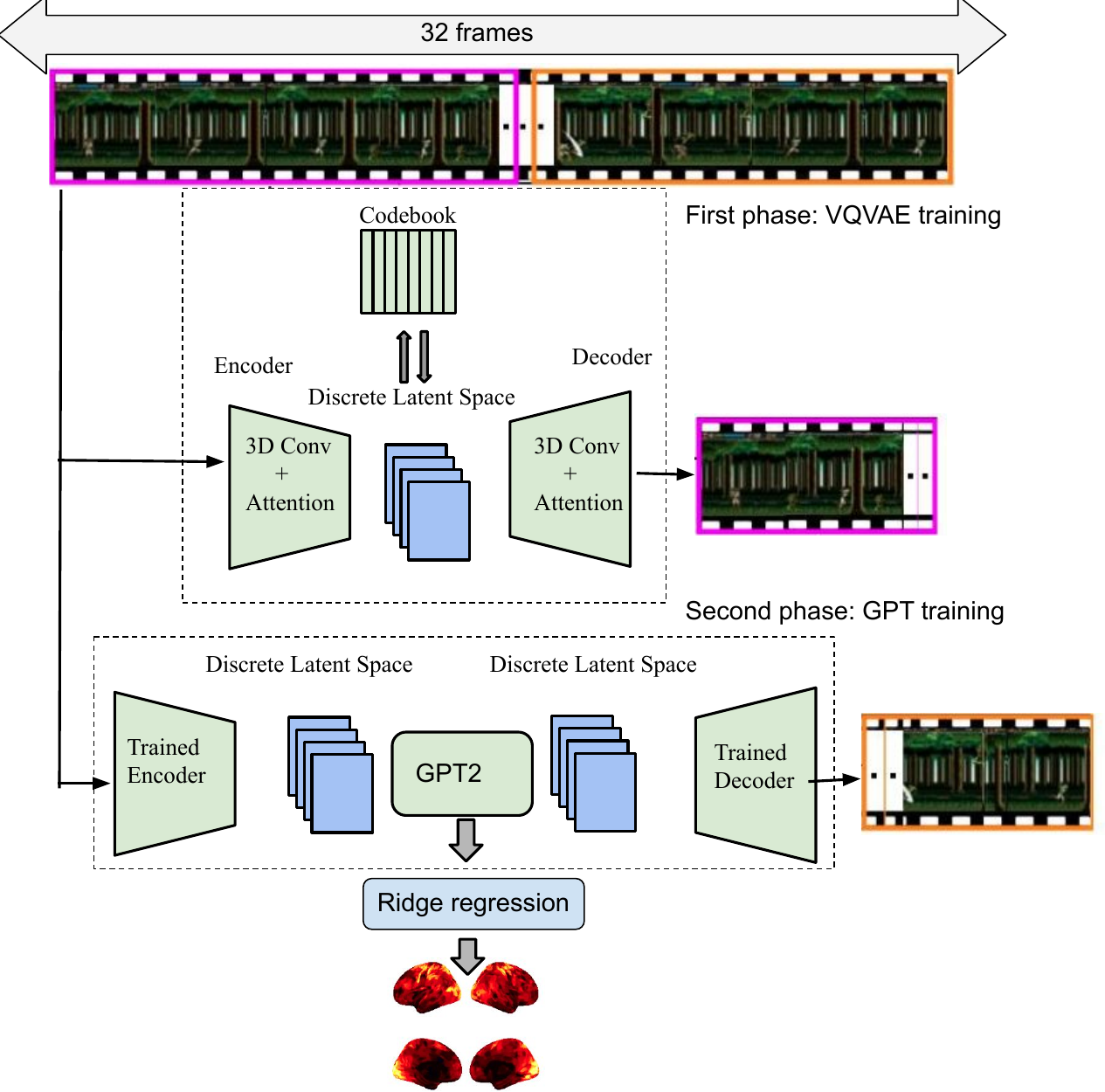}
\caption{Two main steps of brain encoding: Extracting features from movie frames using GPT-2 model and predicting brain response using ridge regression}
        \label{Brain_encoding_steps_shinobi}

\end{figure}

After extracting features from VideoGPT, we applied a 3TR (4.5s) delay to align the features using ridge regression. These final features were then used to train a brain encoding model using ridge regression on pairs of {extracted features, brain activity}. Figure \ref{Brain_encoding_steps_shinobi} illustrates the two main steps of brain encoding: extracting intricate patterns and temporal dependencies in the video sequences and predicting brain responses using ridge regression.

  \subsubsection{VideoGPT model}  

The architecture of the VideoGPT model is an adaptation of VQ-VAE \cite{VQVAE} and GPT-2 \cite{GPT-2_R} architectures. In the first phase of VideoGPT, we trained VQ-VAE to reconstruct 16 sequences of frames. In the second phase, we train GPT to predict the next 16 sequence of frames based on the previous 16 sequences of frames. In the second phased, we use VQ-VAE as a pretrained network to represent sequences of frames with the codebook as the input for GPT.

\subsubsubsection{Training VQ-VAE}
The VQ-VAE \cite{VQVAE} architecture further extends Autoencoders \cite{AE} using discrete latent variables, inspired by vector quantization (VQ). In this approach, the posterior and prior distributions are considered categorical (discrete).  The output of the encoder is compared to all vectors in a (learned) codebook, then the closest codebook vector in the Euclidean distance is selected as input to the decoder. The parameter set of VQ-VAE  includes  parameters of the encoder, decoder, and the embedding space $e$.

In the first phased of Video GPT, a set of discrete latent codes will be trained for the Shinobi dataset through the VQ-VAE, in effect downsampling windows of video frames (sequence length of 16) into a discrete space-time codebook. The encoder of the VQ-VAE will include a series of 3D convolutions followed by attention residual blocks (a replacement for standard residual blocks) to better capture complex spatiotemporal dependencies within video data. In this architecture,  each attention residual block includes Convolution, LayerNorm, position embedding and axial attention layers. The position embedding is shared between all axial attention layers in the encoder and decoder.  The architecture of the decoder starts with attention residual blocks which are then followed by a series of 3D transposed convolution (reverse of encoder) to upsample the video frames across space-time.

\textbf{Multi-Head Attention}: Multi-Head Attention works by projecting the input tensor \( X \) into multiple subspaces using separate attention heads, each computing queries \( Q_i \), keys \( K_i \), and values \( V_i \) with distinct linear transformations. For each head \( i \), the attention is calculated as:

\[
\text{Attention}_i(X) = \text{softmax}\left(\frac{Q_i K_i^T}{\sqrt{d_k}}\right) V_i
\]

where \( d_k \) is the dimensionality of the keys. The outputs from all attention heads are concatenated and projected to obtain the final multi-head attention result:

\[
\text{MultiHeadAttention}(X) = \text{Concat}(\text{Attention}_1(X), \ldots, \text{Attention}_H(X))W^O
\]

This multi-head attention allows the model to process various aspects of the  shinobi video data simultaneously, improving its capacity to learn complex patterns.

\textbf{Axial Attention}:  Axial Attentionis then applied to efficiently manage the high-dimensional video data. Instead of processing the entire tensor at once, axial attention operates along different dimensions separately—width, height, and temporal—allowing the model to focus on specific aspects of the data. Attention is computed separately along each dimension:

- Width Attention:
\[
\text{Attention}_{\text{width}}(X) = \text{softmax}\left(\frac{Q_{\text{width}} K_{\text{width}}^T}{\sqrt{d_k}}\right) V_{\text{width}}
\]

- Height Attention:
\[
\text{Attention}_{\text{height}}(X) = \text{softmax}\left(\frac{Q_{\text{height}} K_{\text{height}}^T}{\sqrt{d_k}}\right) V_{\text{height}}
\]

- Temporal Attention:
\[
\text{Attention}_{\text{temporal}}(X) = \text{softmax}\left(\frac{Q_{\text{temporal}} K_{\text{temporal}}^T}{\sqrt{d_k}}\right) V_{\text{temporal}}
\]

The combined axial attention is represented as:
\[
\text{Attention}_{\text{combined}}(X) = \text{Attention}_{\text{width}}(X) + \text{Attention}_{\text{height}}(X) + \text{Attention}_{\text{temporal}}(X)
\]

This approach allows the model to process different dimensions of the video independently, enhancing its ability to manage complex spatiotemporal structures. Finally, after the attention mechanism, the input tensor is passed through a series of convolutional layers for further feature extraction:

\[
X' = \text{Conv3D}(X)
\]

Axial attention is applied through an axial block:

\[
X'' = \text{AxialBlock}(X')
\]

The output of the axial block is then combined with the original input through a residual connection:

\[
Y = X + X''
\]

This residual connection ensures the retention of the original input features while incorporating the newly learned ones, facilitating smoother training and improving the model’s overall ability to capture both spatial and temporal patterns in video data.

\textbf{Loss Function}:  In VideoGPT, the VQ-VAE is trained using the following loss function where $sg$ refers to a stop-gradient: 
\begin{equation}
L= \norm{x  \minus D(e)} _{2}^{2} +\norm{sg [ \, E(x) ] \, \minus e} _{2}^{2}+ \beta\norm{sg [ \, e ] \, \minus  E(x)  } _{2}^{2}
      \end{equation}

 The loss function includes reconstruction loss $L_{recon}$, a codebook loss $ L_{codebook}$,
and a commitment loss $L_{commit}$. The reconstruction loss controls the VQ-VAE training process to learn efficient representations of frames with minimizing the difference between original and reconstructed frame features. The codebook loss ensures that the codebook embeddings and their corresponding encoder outputs are closely matched based on nearest neighbors lookup.  In the VQ-VAE model, the encoder outputs may oscillate between different code vectors for the same input sample. To tackle this problem, the commitment loss is employed to encourage the encoder outputs to commit to a particular code vector. The commitment loss is weighted by a hyperparameter $\beta$ to  regularise the VQ-VAE training process with penalizes the encoder for switching between code vectors. In the lost function, by stopping the gradients for  $e$ and  pre-trained weights $E(x)$, their values remain fixed. In other words, stop gradient leads to  preserving the representations they have already learned.

\textbf{Optimizer}: The model uses the Adam optimizer (\textit{betas=(0.9, 0.999)}) for training. The model starts with a learning rate of 3e-4, and then the \textit{CosineAnnealingLR} scheduler \cite{CosineAnnealingLR} gradually reduces this learning rate over time according to a cosine function. CosineAnnealingLR scheduler, defined as:

\begin{equation}
\eta_t = \eta_{\text{min}} + \frac{1}{2} (\eta_{\text{max}} - \eta_{\text{min}}) \left(1 + \cos\left(\frac{t}{T_{\text{max}}}\pi\right)\right)
\end{equation}

where \(\eta_{\text{max}}\) is the initial learning rate, and \(T_{\text{max}}\) is the maximum number of training steps. This scheduling allows for a smooth reduction in learning rate, fostering a balance between fast convergence in the initial training phase and fine-tuning as the model nears convergence.

The codebook in the VQ-VAE model is updated using an Exponential Moving Average (EMA) \cite{VQVAE} method to maintain stable and meaningful learned representations. Specifically, the counts \( \text{N} \) and the running average of the embeddings \( \text{z\_avg} \) are updated as follows:

\begin{equation}
\text{N} \leftarrow 0.99 \times \text{N} + 0.01 \times \text{n\_total}
\end{equation}
\begin{equation}
\text{z\_avg} \leftarrow 0.99 \times \text{z\_avg} + 0.01 \times \text{encode\_sum}
\end{equation}

where \( \text{n\_total} \) is the sum of one-hot encoded vectors indicating the frequency of each code being selected, and \( \text{encode\_sum} \) is the sum of the input vectors corresponding to each code. The embeddings are then normalized based on their frequency:

\begin{equation}
\text{weights} = \frac{(\text{N} + 1e-7)}{\left(\sum \text{N}\right) + \text{n\_codes} \times 1e-7} \times \sum \text{N}
\end{equation}
\begin{equation}
\text{embeddings} \leftarrow \frac{\text{z\_avg}}{\text{weights}}
\end{equation}

Unused embeddings (where \( \text{N} < 1 \)) are reinitialized with randomly sampled vectors from the data. This EMA-based update mechanism ensures that the codebook embeddings adapt to new data while remaining stable and representative of the underlying input distribution.

\textbf{Hyperparameters}: As Table \ref{VQVAE_Model_size} shows, the main hyperparameters of the VQ-VAE model include the embedding dimension (\textit{embedding\_dim}), the number of codes in the codebook (\textit{n\_codes}), the number of hidden layers (\textit{n\_hiddens}), the number of residual layers (\textit{n\_res\_layers}), and the downsampling factors for the encoder and upsampling factors for the decoder (\textit{downsample}, \textit{upsample}).

\subsubsubsection{Training autoregressive GPT-2} 
In time series data such as videos, autoregressive models train to predict a time step value (frame in the video) using previous time step values. The VideoGPT autoregressive architecture includes a stack of transformer encoder layers to generate videos (predict next frames) from a latent space. In VideoGPT, through the VQ-VAE (first phase of videoGPT training process), we learn the latent codes, and in the next step, we leverage GPT-2 to model the prior over the latent space. The input to GPT-2 is a sequence of discrete latent codes, produced by the VQ-VAE encoder. The GPT-2 model is trained to generate new sequences of latent codes, which are then passed through the VQ-VAE decoder to produce new video frames.

By employing a self-supervised and autoregressive approach, GPT effectively learns from Shinobi video data, generating meaningful representations of the spatio-temporal dynamics present in the video. This capability enables the model to capture not only the visual information of individual frames but also the temporal transitions that define the sequence. As GPT learns to predict future frames based on previously observed content, it produces contextualized embeddings that reflect the temporal dependencies inherent in video sequences. This rich representation allows the model to discern essential characteristics of movement, object interaction, and scene evolution, making it particularly advantageous for brain encoding, where understanding the flow of time and space of stimuli is crucial.

\textbf{Loss Function}:   The GPT loss function used in the VideoGPT is Cross-Entropy Loss. This loss function is  used for classification tasks and is well-suited for training models like GPT, which predict the next token (or codebook in the context of VQ-VAE) in a sequence.   The loss function measures the difference between the predicted probability distribution and the actual distribution (which is typically a one-hot encoded vector for classification). The formula for the cross-entropy loss is:

\begin{equation}
L_{\text{cross-entropy}} = -\sum_{i=1}^C y_i \log(p_i)
\end{equation}

where \(C\) represents the number of possible latent codes. For each latent code \(i\), \(y_i\) is a one-hot indicator (0 or 1) indicating whether latent code \(i\) is the correct representation for the current input. Additionally, \(p_i\) denotes the predicted probability for latent code \(i\), which is obtained by applying the softmax function to the logits.

\textbf{Optimizer}: Optimization is performed using the Adam optimizer (\textit{betas=(0.9, 0.999)}) with a learning rate of 3e-4. The \textit{CosineAnnealingLR} scheduler adjusts the learning rate according to a cosine function over the training steps, starting at a higher value to enable faster convergence initially, and then decreasing gradually to promote more refined learning as convergence is approached.

\textbf{Hyperparameters}: As Table \ref{GPT2_Model_size} shows, the GPT-2 model employs several hyperparameters, including a hidden dimension (\textit{hidden\_dim}) of 576, 4 attention heads (\textit{heads}), 8 transformer layers (\textit{layers}), and dropout rates of 0.2 for the attention mechanism (\textit{attn\_dropout}) and 0.3 for other layers (\textit{dropout}). 

\begin{table}[t]
\centering
\caption{Model Sizes}
\begin{tabular}{@{}cc@{}}
    \begin{minipage}[t]{0.58\textwidth}
    \scriptsize
    \caption*{(a) VQVAE Model Size}
    \renewcommand{\arraystretch}{1.2} 
    \begin{tabular}{|c|c|c|}
        \hline
        Name & Type & Parameters \\
        \hline \hline
        encoder & encoder & 18.7 M \\
        \hline
        decoder & decoder & 17.1 M \\
        \hline
        pre\_vq\_conv & SamePadConv3d & 30.8 K \\
        \hline
        post\_vq\_conv & SamePadConv3d & 31.0 K \\
        \hline
        codebook & Codebook & 0 \\
        \hline \hline
        \multicolumn{2}{|c|}{Trainable params} & 35.8 M \\
        \hline
        \multicolumn{2}{|c|}{Non-trainable params} & 0 \\
        \hline
        \multicolumn{2}{|c|}{Total params} & 35.8 M \\
        \hline
        \multicolumn{2}{|c|}{Total model params size (MB)} & 143.306 \\
        \hline
    \end{tabular}
    \label{VQVAE_Model_size}
    \end{minipage}
    &
    \begin{minipage}[t]{0.47\textwidth}
    \centering
    \scriptsize
    \caption*{(b) GPT-2 Model Size}
    \renewcommand{\arraystretch}{1.2} 
    \begin{tabular}{|c|c|c|}
        \hline
        Name & Type & Parameters \\
        \hline \hline
        vqvae & VQ-VAE & 35.8 M \\
        \hline
        resnet & ResidualBlock & 2.4 M \\
        \hline
        fc\_in & Linear & 73.7 K \\
        \hline
        attn\_stack & AttentionStack & 37.2 M \\
        \hline
        fc\_out & Linear & 589 K \\
        \hline \hline
        \multicolumn{2}{|c|}{Trainable params} & 40.3 M \\
        \hline
        \multicolumn{2}{|c|}{Non-trainable params} & 35.8 M \\
        \hline
        \multicolumn{2}{|c|}{Total params} & 76.2 M \\
        \hline
        \multicolumn{2}{|c|}{Total model params size (MB)} & 304.606 \\
        \hline
    \end{tabular}
    \label{GPT2_Model_size}
    \end{minipage}
\end{tabular}

\vspace{1cm}

\begin{minipage}[t]{0.9\textwidth}
\centering
\scriptsize
\caption*{(c) Hyperparameters for VQVAE and GPT-2 components}
\renewcommand{\arraystretch}{1.2} 
\begin{tabular}{|c|c|c|}
    \hline
    \textbf{Parameter} & \textbf{VQVAE} & \textbf{GPT-2} \\
    \hline \hline
    embedding\_dim & 256 & - \\
    \hline
    n\_codes & 2048 & - \\
    \hline
    n\_hiddens & 240 & - \\
    \hline
    n\_res\_layers & 4 & - \\
    \hline
    downsample & (4, 4, 4) & - \\
    \hline
    hidden\_dim & - & 576 \\
    \hline
    heads & 2 & 4 \\
    \hline
    layers & - & 8 \\
    \hline
    dropout & - & 0.2 \\
    \hline
    attn\_dropout & - & 0.3 \\
    \hline
\end{tabular}
\label{VQVAE_VideoGPT_Hyperparameters}
\end{minipage}

\end{table}
\subsubsection{Ridge regression}
\label{sec:ridge}
Ridge regression was first proposed by Hoerl and Kennard~\cite{Ridge1} as a generalization of ordinary least-square regression. In comparison with ordinary least mean square regression, ridge regression provides better generalization to unseen data through regularization of coefficient estimates, in particular in the presence of a large number of predictor variables. The ridge regression is expressed as the following optimization problem solving for regression coefficients $b^{\ast}$ independently at each spatial location:
\begin{equation}
     b^{\ast} = \argmin_{b \in \mathbb{R} ^{p}} \left( \Vert y - Xb \Vert_{2} ^ {2} +  \lambda  \Vert b\Vert _{2} ^{2}\right), \label{eq:ridge}
\end{equation}
where $X \in \mathbb{R} ^{n \times p}$ is the matrix of stimuli features with $n$ time samples and $p$ features, $\Vert .\Vert _{2}$ is the $ \boldsymbol\ell^2$ norm of a vector, and $y \in \mathbb{R} ^ n$ is the target vector obtained from fMRI data at a single spatial location. The hyper-parameter $\lambda$ is used to control the weighting of the penalty in the loss function. The best value for $\lambda$ is estimated among a set of candidate values through cross-validation, as explained below. If the value of $\lambda$ is too low the training process may overfit and if the value of $\lambda$ is too high then the brain encoder model may underfit~\cite{banded1}. 

\subsubsection{Brain encoding performance and hyper-parameter optimization}
For a given subject, the samples $X$ were split into training (90\% random) and test (10\% remaining) subsets. The coefficients of the ridge regression were selected through Eq. \ref{eq:ridge} based on the training set only. We measured the final quality of brain encoding as the Pearson's correlation coefficient between the actual fMRI time series and the time series predicted by the ridge regression model, on the test set. A leave-one-out validation was used inside the training set to estimate the hyper-parameter value $\lambda$ with optimal performance (based on cost function defined in Eq. \ref{eq:ridge}), based on the grid:
$$\lambda \in \left\{0.1, 1, 100 \right\}.$$

\subsubsection{Computational environment}
Brain encoding experiments were run on Beluga, a high-performance computing (HPC) cluster of Canada Digital Alliance, providing researchers with a robust infrastructure for advanced scientific computations. Beluga features numerous compute nodes, high-speed interconnects, and parallel processing capabilities, visit the \href{https://docs.alliancecan.ca/mediawiki/index.php?title=B%C3%A9luga/en} {Beluga technical documentation} page for details. 
\subsection {Scaling up VideoGPT model}
\subsubsection{Scaling up GPT in terms of Dataset Size}
The GPT model was trained on datasets of varying sizes: 10k, 100k, 1M, and 6M samples. These datasets were derived from the entire available dataset collected from four subjects who played the Shinobi game while the videos were recorded. After training the GPT model on these different dataset sizes, we extracted activations from the model and compared the brain encoding results. In all scenarios of varying dataset size, the hyperparameters related to model size were fixed according to Table \ref{VQVAE_VideoGPT_Hyperparameters}.

\subsubsection{Scaling up GPT in terms of Model Size}

In this study, the scaling of the GPT model size was explored by varying the following hyperparameters:

\begin{itemize}
    \item \textbf{Number of hidden layers}: 1, 2, 4, 8
    \item \textbf{Dimension of hidden layers}: 6, 15, 30, 63, 126, 255, 576, 1024 (note: the hidden dimension must be a multiple of 3 and greater than 3)
    \item \textbf{Number of heads in multi-head attention mechanism}: 1, 2, 4, 8
\end{itemize}

In each scenario, the other model hyperparameters, as detailed in Table \ref{VQVAE_VideoGPT_Hyperparameters}, were kept fixed. Furthermore, the GPT model was trained on a dataset of 6 million samples across all scenarios. After training with these configurations, activations were extracted from the GPT model, and the brain encoding results were compared using actual fMRI data. Specifically, we used the \text{"attn\_stack.attn\_nets.4.post\_fc\_dp"} layer to represent the GPT activations for experiments involving the hidden dimension and the number of heads, but only for the case of the number of layers. Table \ref{tab:NUmber of layers_brain_encoding_layers} shows the selected layer for brain encoding:

\begin{table}[h]
\small
\centering
\caption{Selected GPT layers for brain encoding based on the number of layers}
\label{tab:NUmber of layers_brain_encoding_layers}
\begin{tabular}{|c|l|}
\hline
Number of Layers & Selected GPT Layer for Brain Encoding \\ \hline
1 & attn\_stack.attn\_nets.0.post\_fc\_dp \\ \hline
2 & attn\_stack.attn\_nets.1.post\_fc\_dp \\ \hline
4 & attn\_stack.attn\_nets.3.post\_fc\_dp \\ \hline
8 & attn\_stack.attn\_nets.7.post\_fc\_dp \\ \hline
\end{tabular}
\end{table}

\subsubsection{High Performance computing during GPT training process} 
In \cite{B8}, the authors proposed an approach for training deep neural networks based on half-precision floating-point numbers, without losing model accuracy or modifying the hyperparameters. In this work, the results show that the mixed-precision approach has two main advantages: 1) approximately
halves the memory requirements, 2) accelerating arithmetic on GPU. In the proposed approach
weights, activations, and gradients are stored in half-precision format. Updating the parameters contains two key steps: Firstly, a single-precision copy of weights is maintained to accumulate the
gradients after backpropagation. Secondly, scaling the loss function to preserve gradient values with small magnitudes. The mixed precision approach works across a wide variety of modern large scale Deep Learning model architectures, trained on large datasets.

 Two commonly used 16-bit formats are float 16 and bfloat16, each with distinct advantages. Float 16, adhering to the IEEE 754 standard, uses 1 bit for the sign, 5 bits for the exponent, and 10 bits for the mantissa. Its compact nature reduces memory usage and accelerates computations but limits precision and dynamic range. It is widely used in scenarios where speed and memory efficiency are critical, though it struggles with representing very large or small numbers. In contrast, bfloat16 allocates 1 bit for the sign, 8 bits for the exponent, and 7 bits for the mantissa, offering the same dynamic range as float 32 but with reduced precision. This format is optimized for modern hardware, including Google’s TPUs and NVIDIA GPUs such as A100, making it a popular choice in machine learning.

Recently, the FP8 format has emerged as a breakthrough for further improving memory and computational efficiency. With NVIDIA's H100 (Hopper architecture), two new FP8 formats are introduced:
\begin{itemize}
     \item E5M2 (5 bits for the exponent, 2 bits for the mantissa).
    \item E4M3 (4 bits for the exponent, 3 bits for the mantissa).
\end{itemize}

These formats strike a balance between performance and precision, allowing for even faster computations and more efficient memory usage, making them ideal for deep learning workloads. Hopper’s dynamic precision capabilities ensure seamless switching between FP8, FP16, and FP32, adapting to the specific needs of the task at hand and maximizing overall efficiency. 

In this work, we focus on comparing the training of GPT-2 using 32-bit and standard 16-bit precision format, and we discuss the impacts of these precision levels on brain encoding.

\begin{comment}  
\subsubsection{Distributed Data parallelism}

  \begin{figure}[t]
\centering
\includegraphics[width=9cm]{dataparralelism_new.png}
\caption{Distributed data parallelism of training VideoGPT for shinobi dataset}
        \label{ddp}

\end{figure} 

\end{comment}

\section{Results}

\begin{figure}[tp]
\centering

% First row with figures a and b
\begin{subfigure}[b]{0.41\textwidth}
    \centering
    \includegraphics[width=\textwidth]{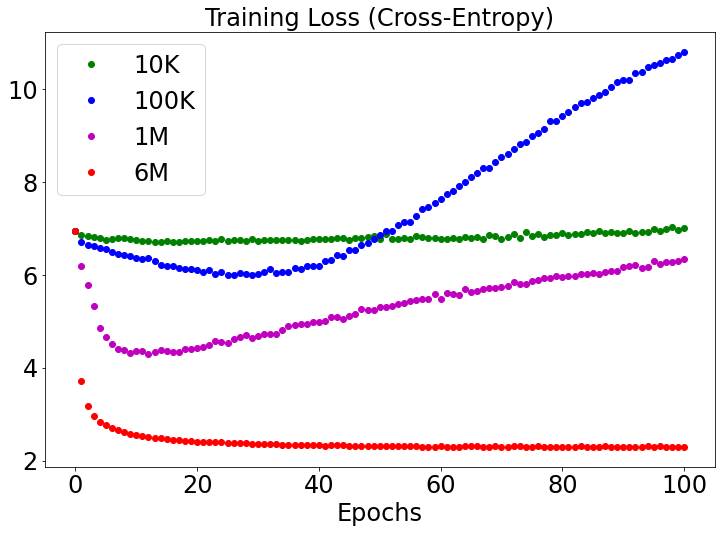}
    \caption{GPT training loss}
    \label{fig:GPT_training_loss_data_size}
\end{subfigure}

\hspace{1\textwidth}
\begin{subfigure}[b]{0.61\textwidth}
    \centering
    \includegraphics[width=\textwidth]{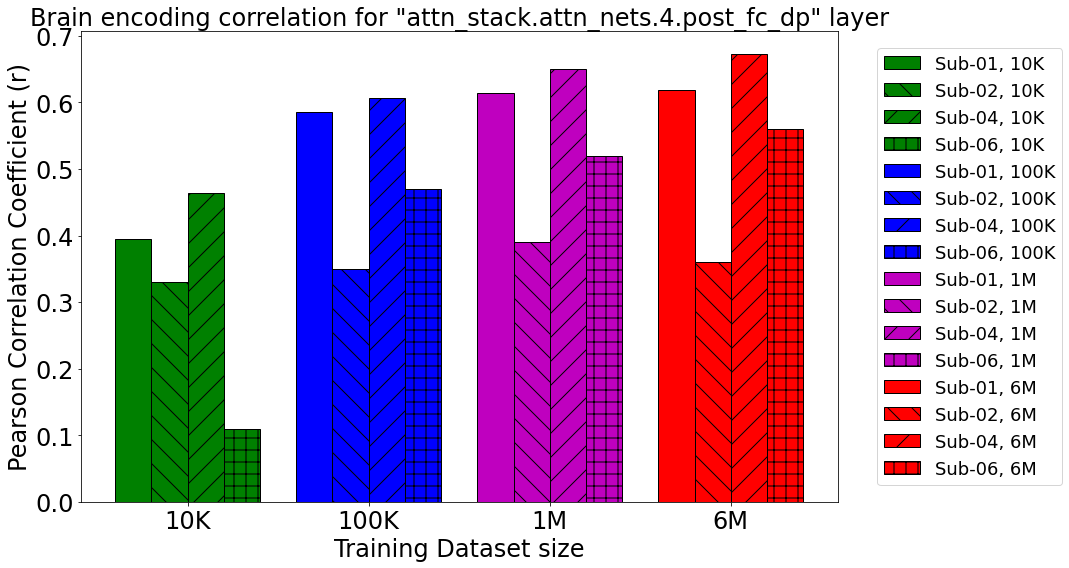}
    \caption{Brain encoding correlations across subjects}
    \label{fig:Brain_encoding_correlations_data_size}
\end{subfigure}

\vspace{0.3cm}

% Second row with larger figure c
\begin{subfigure}[b]{0.65\textwidth}
    \centering
    \includegraphics[width=\textwidth]{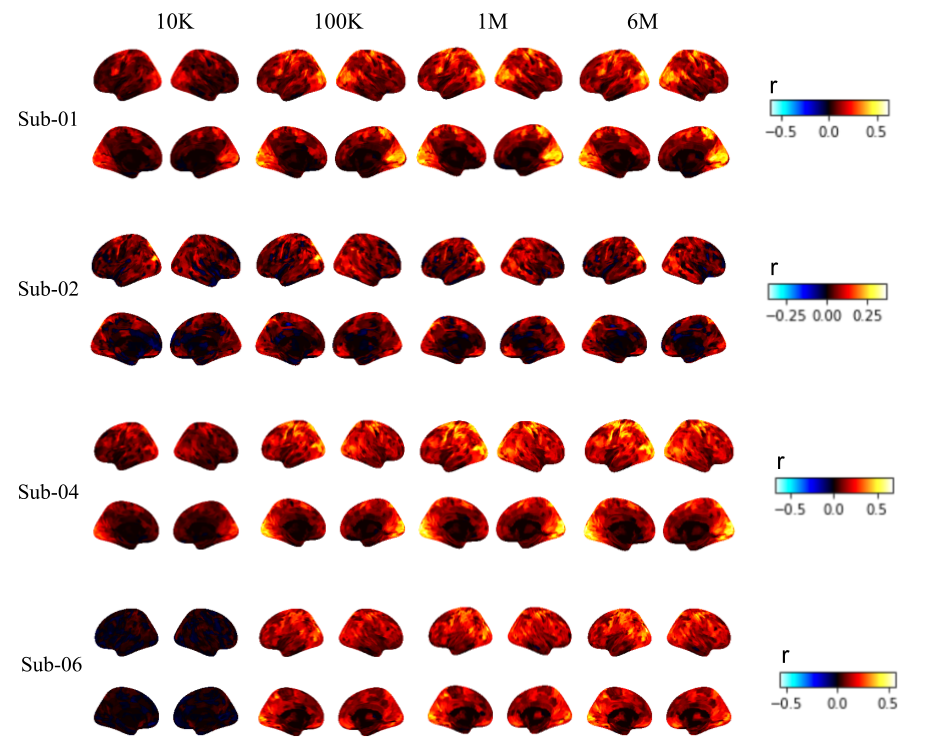}
    \caption{Brain maps across subjects}
    \label{fig:Brain_maps_data_size}
\end{subfigure}
\caption{GPT training across different training dataset
sizes}
\end{figure}

\subsection{Scaling up the dataset size for training VideoGPT resulted in a significant improvement in brain encoding performance} 
In this work, we explored the impact of varying training dataset sizes  of GPT model in brain encoding tasks. Specifically, we trained the GPT model on datasets of sizes 10K, 100K, 1M, and 6M samples, derived from video recordings of subjects playing the Shinobi game.

The training loss decreases more rapidly and stabilizes at a lower value with larger datasets, as shown in Figure \ref{fig:GPT_training_loss_data_size}. Specifically, GPT trained on the 6M dataset (red curve) shows a rapid decline in cross-entropy loss, reaching a stable point around epoch 40. In contrast, the models trained on smaller datasets, such as the 10K (green curve) and 100K (blue curve) datasets, exhibit slower convergence and higher final loss values, with the 100K dataset even experiencing an upward trend after epoch 60, indicating overfitting. The model trained on the 1M dataset (purple curve) falls in between these extremes, achieving better performance than the smaller datasets but not reaching the low loss values of the 6M dataset. This trend demonstrates that increasing the dataset size leads to better GPT convergence, with the larger datasets enabling the model to generalize better and achieve a lower cross-entropy loss, thus improving overall training efficiency and effectiveness.

Figure \ref{fig:Brain_encoding_correlations_data_size}  demonstrates the impact of training dataset size on brain encoding correlations, as indicated by the Pearson correlation coefficient (r) between actual and predicted brain activity across various subjects. The GPT model trained on the 6M dataset (red bars) generally achieves the highest correlation values across all subjects, highlighting the enhanced brain encoding predictive accuracy with increased GPT training data. 

The brain encoding results of Subject 01 show a noticeable improvement in correlation from approximately 0.4 with the 10K dataset to nearly 0.6 with the 6M dataset, representing an increase by a factor of about 1.5. Similarly, Subject 06 exhibits a significant enhancement, with correlations rising from around 0.15 with the 10K dataset to almost 0.55 with the 6M dataset, leading to an improvement by a factor of approximately 3.7. On the other hand, Subject 02 shows a modest improvement, where the correlation increases from around 0.35 with the 10K dataset to about 0.4 with the 6M dataset, yielding only a 1.14 times improvement.

This pattern suggests that while larger training datasets generally lead to more accurate brain encoding, the extent of improvement varies across subjects. For example, Sub-06 experienced significant gains, while Sub-02 saw less substantial improvements, indicating that the effectiveness of dataset scaling may differ depending on individual subject characteristics. Figure \ref{fig:Brain_maps_data_size} illustrates brain maps across four subjects.

\subsection{Scaling up the GPT model size in terms of hidden dimensions resulted in a significant improvement in brain encoding performance}

\begin{figure}[tp]
\centering

% First row with figures a and b
\begin{subfigure}[b]{0.57\textwidth} % Adjusted width
    \centering
    \includegraphics[width=\textwidth]{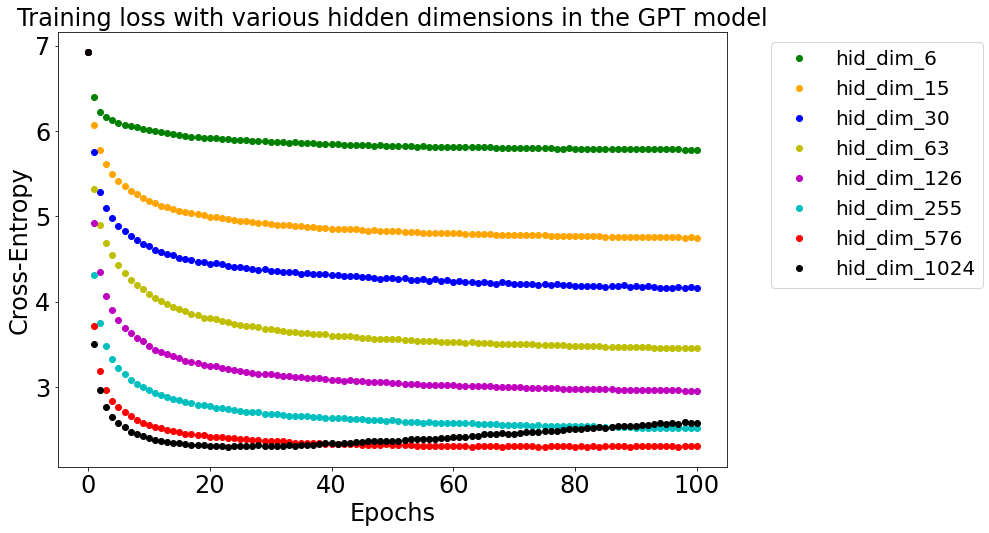}
    \caption{\scriptsize GPT training loss}
    \label{fig:GPT_training_loss_hidd}
\end{subfigure}

\vspace{0.4cm}

\begin{subfigure}[b]{0.67\textwidth} % Adjusted width
    \centering
    \includegraphics[width=\textwidth]{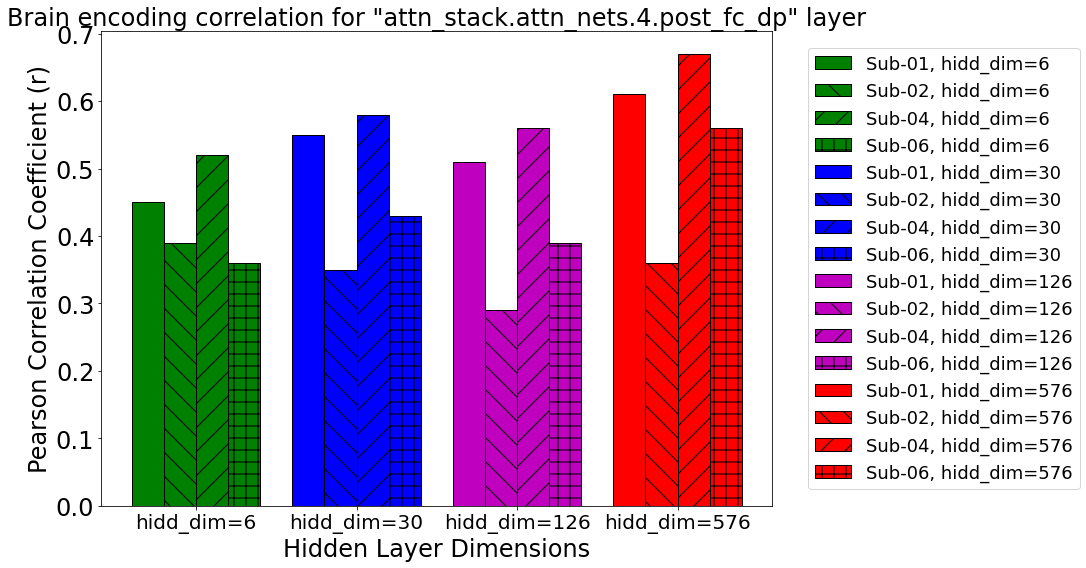}
    \caption{\scriptsize Brain encoding correlation coefficients across subjects}
    \label{fig:Brain_encoding_correlations_hidd}
\end{subfigure}

\vspace{0.4cm}

% Second row with larger figure c
\begin{subfigure}[b]{0.65\textwidth}
    \centering
    \includegraphics[width=\textwidth]{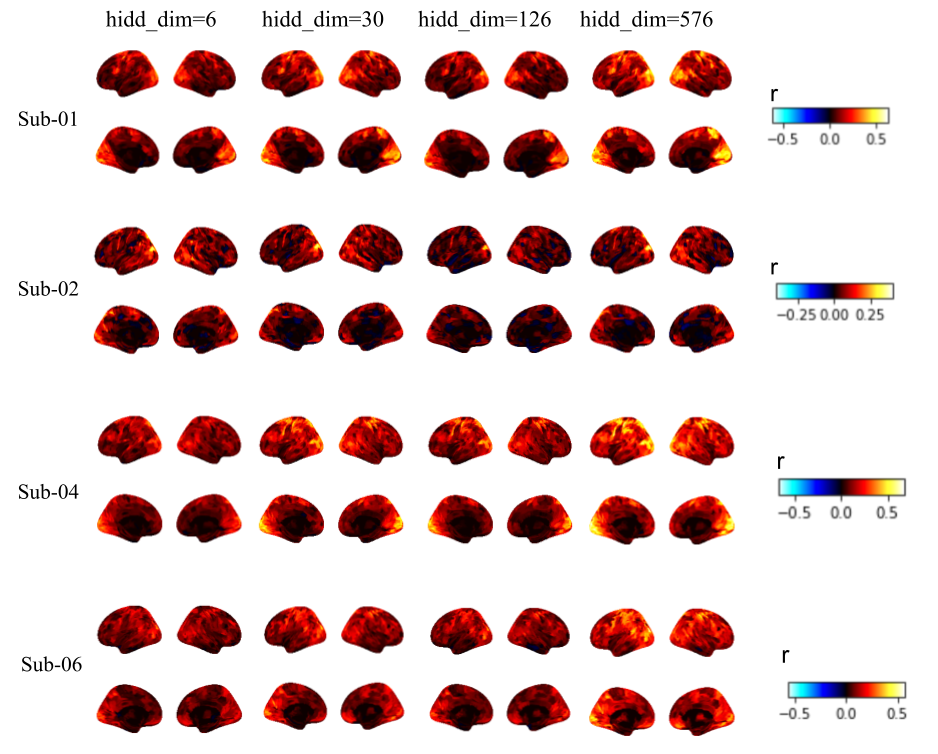}
    \caption{\scriptsize Brain maps across subjects}
    \label{fig:Brain_maps_hidd}
\end{subfigure}
\caption{ GPT training across different hidden dimensions}
\end{figure}

In this section, we investigated the impact of increasing the hidden dimensions in the GPT model on brain encoding performance. The VideoGPT model was trained with various hidden dimensions, while other hyperparameters were kept constant, as outlined in Table \ref{VQVAE_VideoGPT_Hyperparameters}.

Figure \ref{fig:GPT_training_loss_hidd} displays the training loss curves across different hidden dimensions. As the hidden dimensions increase from 6 to 576, the training loss decreases more rapidly and stabilizes at lower cross-entropy values. This indicates that models with larger hidden dimensions have greater capacity, leading to better convergence during training. However, after epoch 30, the loss for the model with 1024 hidden dimensions begins to increase slightly, suggesting overfitting. Then, the results hint at a slight decline in performance beyond a certain threshold, as evidenced by the small drop in correlation for some subjects at the highest hidden dimension tested (1024).

Figure \ref{fig:Brain_encoding_correlations_hidd} shows the Pearson correlation coefficients between the model predictions and actual brain activity across different subjects. The results demonstrate that models with larger hidden dimensions generally achieve higher brain encoding correlations. Notably, at a hidden dimension of 576, the highest correlations are observed, particularly for Subjects 01, 04, and 06, where the improvements are significant compared to the hidden dimension of 126.

However, the pattern is not entirely consistent across all subjects and hidden dimensions. For instance, at a hidden dimension of 30, Subject 02 exhibits a decrease in brain encoding performance compared to its performance at a hidden dimension of 6, while other subjects show an increase. Furthermore, at a hidden dimension of 126, all subjects experience a decrease in correlation compared to the results at 30. These fluctuations highlight the nuanced relationship between model complexity and brain encoding accuracy. Figure \ref{fig:Brain_maps_hidd} presents brain activation maps for four subjects across different hidden dimensions.

\subsection{Effect of increasing layers of GPT on brain encoding performance is limited}

In this section, we explored the effect of varying the number of layers in the GPT model on brain encoding performance. The GPT model was trained with 1, 2, 4, 8 and 16 layers, while all other hyperparameters were kept constant, as detailed in Table \ref{VQVAE_VideoGPT_Hyperparameters}.  

\begin{figure}[tp]
\centering

% First row with figures a and b
\begin{subfigure}[b]{0.5\textwidth} % Adjusted width
    \centering
   \includegraphics[width=\textwidth]{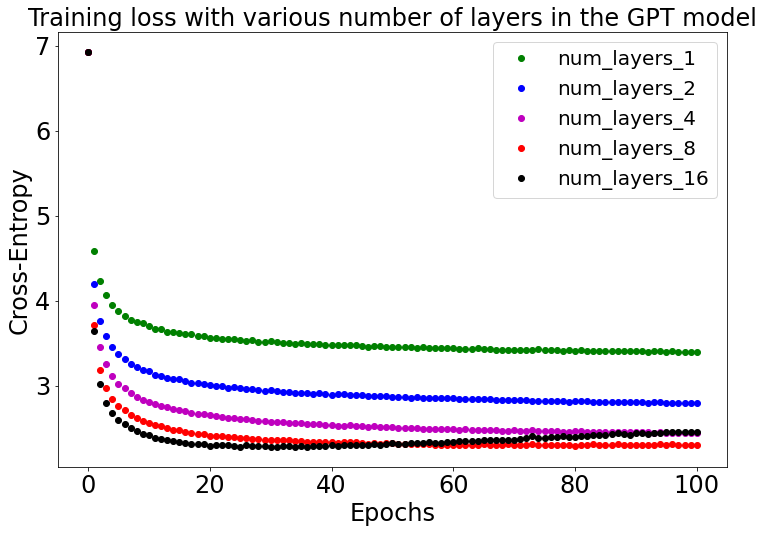}
    \caption{\scriptsize GPT training loss}
    \label{fig:GPT_training_loss_num_layers}
\end{subfigure}

\vspace{0.4cm}

\begin{subfigure}[b]{0.6\textwidth} % Adjusted width
    \centering
     \includegraphics[width=\textwidth]{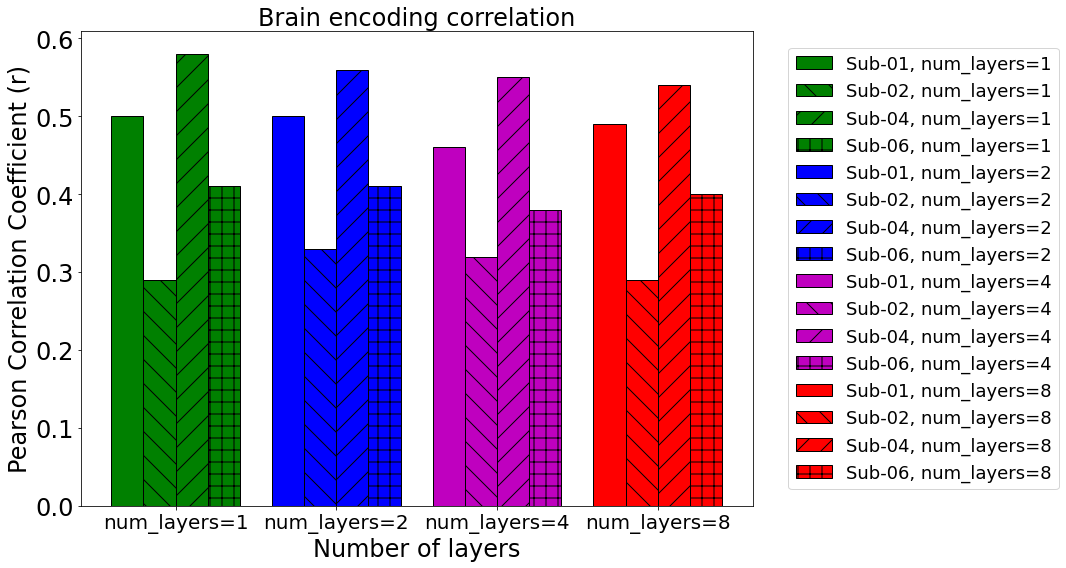}
    \caption{\scriptsize Brain encoding correlation coefficients across subjects}
    \label{fig:Brain_encoding_correlations_num_layers}
\end{subfigure}

\vspace{0.4cm}

% Second row with larger figure c
\begin{subfigure}[b]{0.65\textwidth}
    \centering
    \includegraphics[width=\textwidth]{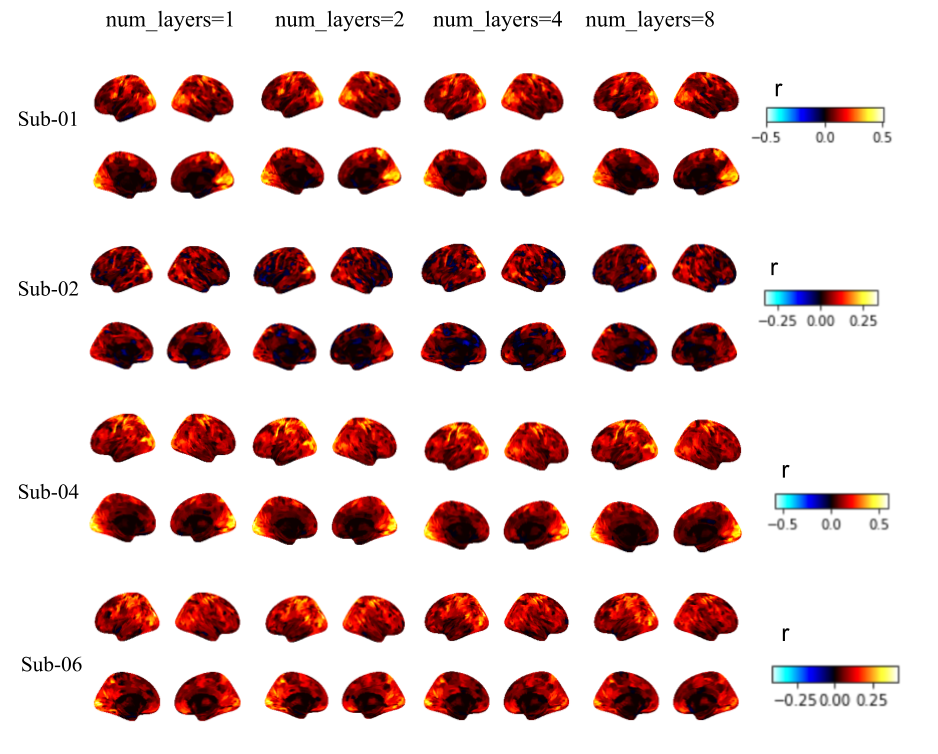}
    \caption{\scriptsize Brain maps across subjects}
    \label{fig:Brain_maps_num_layers}
\end{subfigure}
 \caption{GPT training across different number of layers}
\end{figure}

Figure \ref{fig:GPT_training_loss_num_layers} illustrates the training loss curves across different layer configurations. As the number of layers increases from 1 to 8, we observe a notable decrease in the training loss, particularly during the early epochs. This indicates that deeper models have a greater capacity for capturing complex patterns in the data, leading to more effective convergence. However, beyond 4 layers, the rate of improvement in the training loss begins to plateau, suggesting that the benefits of additional layers reduce as the model becomes deeper. Additionally, the model with 16 layers shows signs of overfitting after epoch 30, as indicated by a minor increase in the training loss, implying that excessively deep models may lead to overfitting on the training data.

Figure \ref{fig:Brain_encoding_correlations_num_layers} presents the Pearson correlation coefficients between the model predictions and actual brain activity across different subjects (sub-01, sub-02, sub-04, and sub-06) for varying numbers of layers (1, 2, 4, and 8). The corresponding GPT layers selected for brain encoding in each scenario are listed in Table \ref{tab:NUmber of layers_brain_encoding_layers}.  The results show that models with 1 layer generally perform well, with sub-01, sub-04 and sub-06,  achieving the highest correlation. When increasing the number of layers to 2,  there is an observable improvement in performance across  Sub-02. The correlation coefficients do not show consistent improvement with increases in number of layers.  This suggests that  there is a point where  increasing  of GPT model layer depth does not yield better performance of brain encoding and may even result in diminishing in correlations.  Figure \ref{fig:Brain_maps_num_layers} illustrate the brain maps across six subjects.

\subsection{Increasing the number of attention heads  does not  translate to better brain encoding performance}
\begin{figure}[tp]
\centering

% First row with figures a and b
\begin{subfigure}[b]{0.45\textwidth} % Adjusted width
    \centering
   \includegraphics[width=\textwidth]{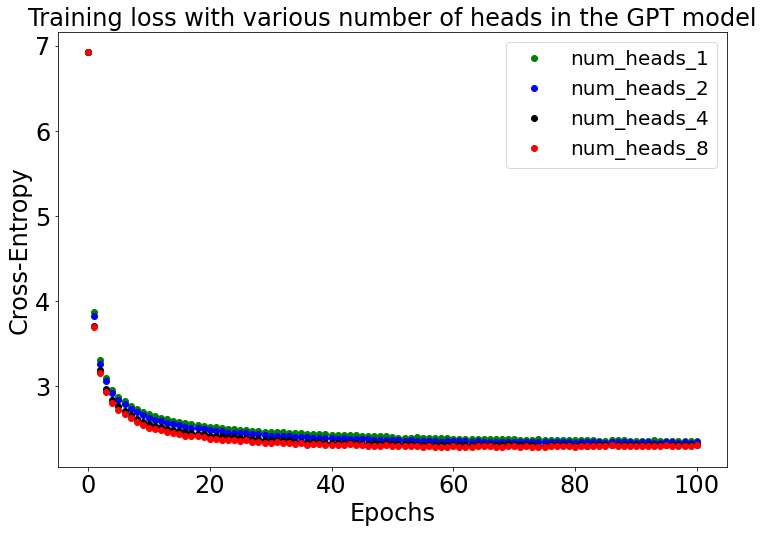}
    \caption{\scriptsize GPT training loss}
    \label{fig:GPT_training_loss_head_num}
\end{subfigure}

\vspace{0.4cm}

\begin{subfigure}[b]{0.6\textwidth} % Adjusted width
    \centering
     \includegraphics[width=\textwidth]{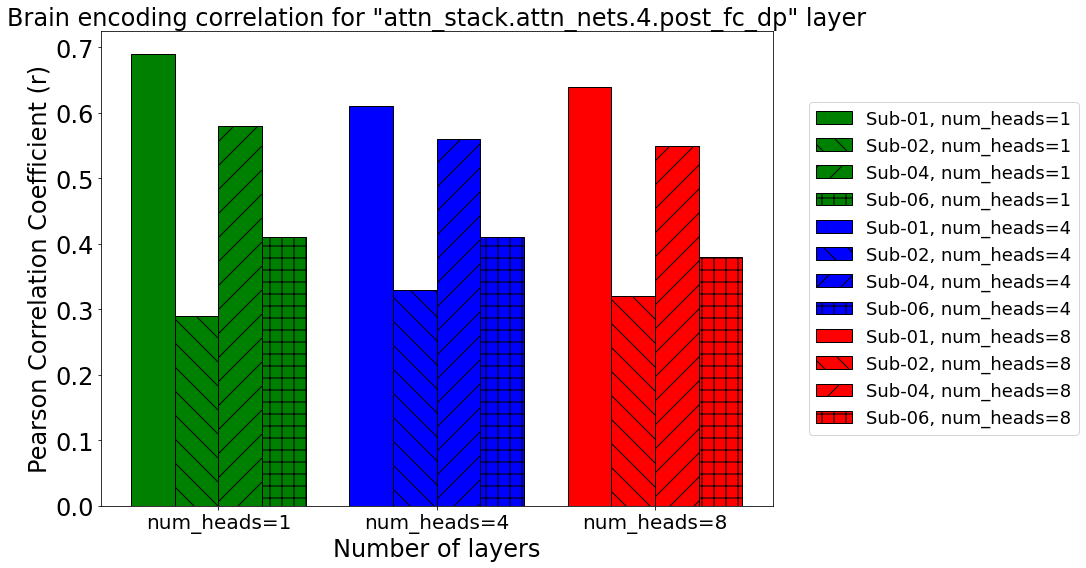}
    \caption{\scriptsize Brain encoding correlation coefficients across subjects}
    \label{fig:Brain_encoding_correlations_head_num}
\end{subfigure}

\vspace{0.4cm}

% Second row with larger figure c
\begin{subfigure}[b]{0.60\textwidth}
    \centering
    \includegraphics[width=\textwidth]{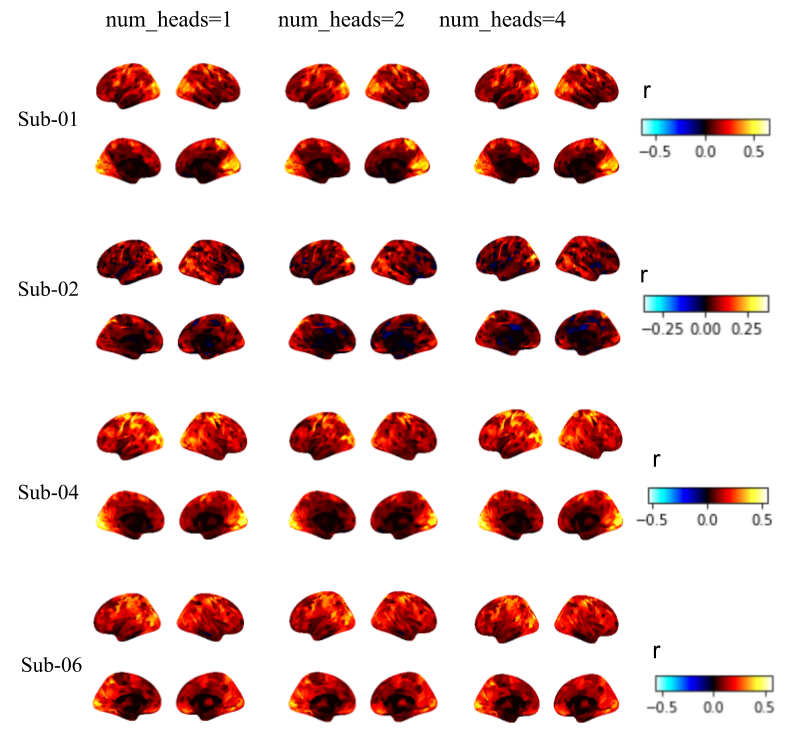}
    \caption{\scriptsize Brain maps across subjects}
    \label{fig:Brain_maps_head_num}
    
\end{subfigure}
\caption{Training GPT with FP32 and FP16 precision}
\end{figure}
In this section, we investigated the effect of changing the number of attention heads in the GPT model on brain encoding performance. The model was trained with varying numbers of attention heads (1, 4, and 8), while other hyperparameters were kept constant in Table \ref{VQVAE_VideoGPT_Hyperparameters}.

Figure \ref{fig:GPT_training_loss_head_num} shows the training loss curves across different numbers of attention heads. The training loss decreases steadily and converges similarly across all configurations, regardless of the number of heads. However, there is no significant difference in the rate of convergence or the final cross-entropy values, suggesting that the number of attention heads does not have impact on training of GPT efficiency.

Figure \ref{fig:Brain_encoding_correlations_head_num} presents the Pearson correlation coefficients between model predictions and actual brain activity across different subjects. The results indicate that brain encoding performance is somewhat sensitive to the number of attention heads. Notably, models with 1 attention head exhibit noticeable higher correlation values for Sub-01 compared to models with 4 and 8 attention heads. However, for Sub-02, models with 2 attention heads show slightly better performance compared to models with 4 and 8 heads. For Sub-04 and Sub-06, varying the number of attention heads does not significantly affect brain encoding performance. Increasing the number of attention heads  does not always translate to better brain encoding performance.

\subsection{Training GPT with 32-bit and 16-bit floating-point precision yields exactly the same results for brain encoding}

\begin{figure}[tp]
\centering

% First row with figures a and b
\begin{subfigure}[b]{0.4\textwidth} % Adjusted width
    \centering
    \includegraphics[width=\textwidth]{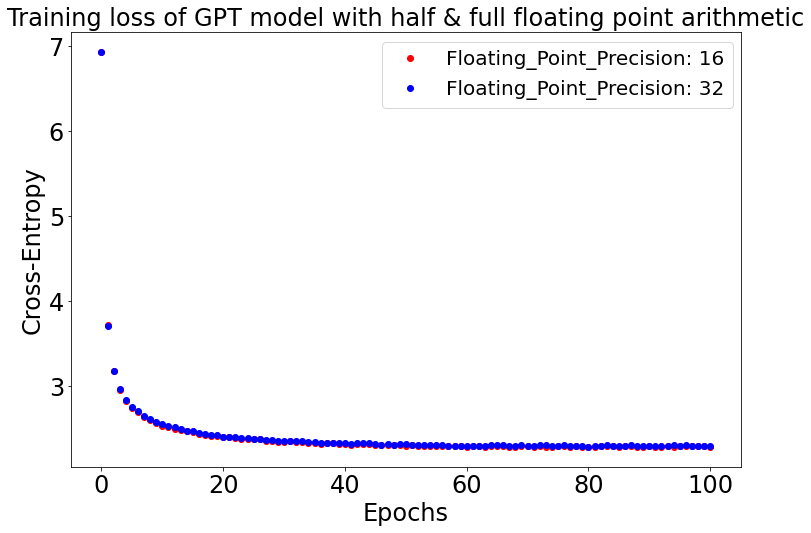}
    \caption{\scriptsize GPT training loss}
    \label{fig:GPT_training_loss_FP}
\end{subfigure}

\vspace{0.4cm}

\begin{subfigure}[b]{0.65\textwidth} % Adjusted width
    \centering
    \includegraphics[width=\textwidth]{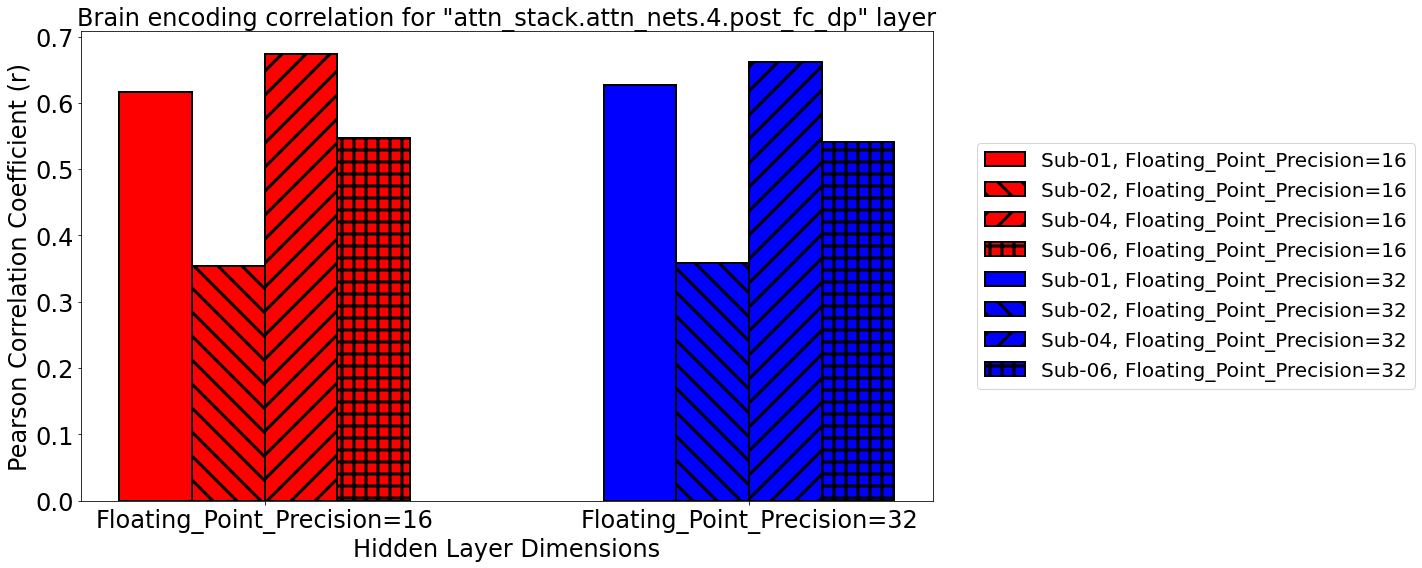}
    \caption{\scriptsize Brain encoding correlation coefficients across subjects}
    \label{fig:Brain_encoding_correlations_FP}
\end{subfigure}

\vspace{0.4cm}

% Second row with larger figure c
\begin{subfigure}[b]{0.55\textwidth}
    \centering
    \includegraphics[width=\textwidth]{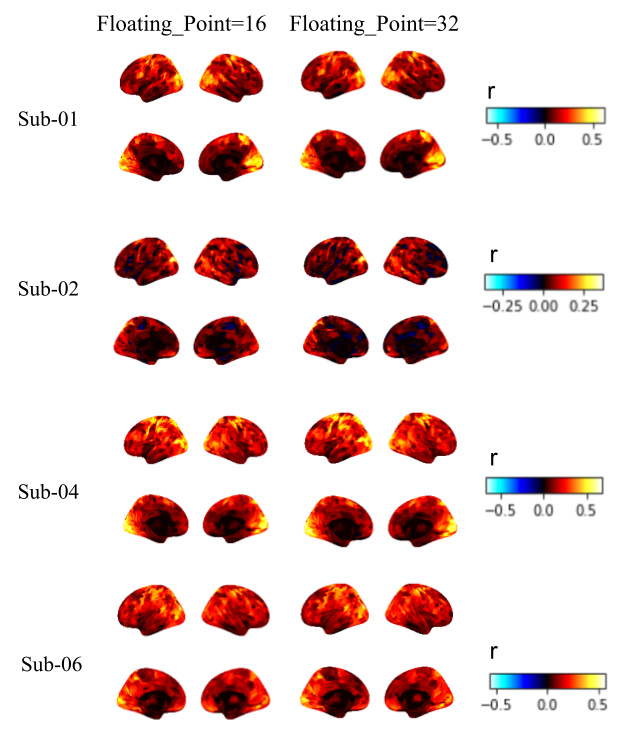}
    \caption{\scriptsize Brain maps across across subjects}
    \label{fig:Brain_maps_FP}
\end{subfigure}
\caption{Training GPT with FP32 and FP16 precision}
\end{figure}

The training time of GPT with 32-bit floating-point precision was 35 hours, while training with 16-bit floating-point precision took 30 hours. The computational environment included 5 CPUs and 4x Tesla V100-SXM2-16GB GPUs in the Beluga HPC computing cluster. This means that using standard 16-bit precision sped up training by ~ 1.17 times. 

For this experiment, we leveraged PyTorch Lightning framework to enable mixed precision GPT training. The model size of GPT is presented in  Table.\ref{GPT2_Model_size}. We utilized the standard FP16 format, which consists of 1 sign bit, 5 exponent bits, and 10 mantissa bits. The Beluga dashboard provided execution time and memory statistics, indicating that 16-bit precision was more efficient. 

As shown in Figure \ref{fig:GPT_training_loss_FP}, training GPT with both 32-bit and 16-bit precision results in the same loss function behavior over epochs, with identical convergence rates. We then benchmarked the effects on brain encoding and found that the results were exactly the same  for all six subjects, as depicted in Figure \ref{fig:Brain_encoding_correlations_FP}. Thus, training with 16-bit floating-point precision (standard FP16 format) leads to  1.17 times speed-up without any impact on brain encoding prediction accuracy. The loss values for training GPT were identical in both scenarios. In terms of brain encoding results, the brain map regions were compatible, and our findings indicated that the maximum correlation values were the same in both scenarios.

\section{Conclusion}

In this study, we assessed the influence of scaling dataset size, hidden dimensions, layers, attention heads, and floating-point precision on the performance of the VideoGPT model in brain encoding tasks. The findings reveal that increasing the training dataset size leads to substantial improvements in brain encoding performance, as evidenced by lower training losses and higher Pearson correlation coefficients across subjects. Notably, models trained on larger datasets enhanced predictive accuracy, underscoring the importance of data quantity for effective brain encoding. Similarly, scaling the model's hidden dimensions improved performance, with an optimal increase observed at 576 hidden dimensions, while further increases resulted in diminishing returns and potential overfitting. Conversely, the number of layers had a limited effect on performance, with noticeable gains up to eight layers, beyond which the improvements plateaued and overfitting became evident. The investigation into the number of attention heads revealed that changes in this parameter had little impact on overall performance, indicating that the effectiveness of attention mechanisms may not directly correlate with their quantity. Finally, training GPT with both 32-bit and 16-bit floating-point precision yielded identical results, indicating that the choice of precision does not significantly impact the model's performance in this context. This finding enables us to utilize mixed precision training to accelerate the training time of transformer models on stimuli without sacrificing brain encoding correlations. Overall, the results highlight the critical role of dataset size and model complexity in enhancing brain encoding capabilities, providing insights for future research and applications in neural decoding and cognitive modeling.

\section{Availability of code and data}
The code to reproduce our experiments is available at \url{https://github.com/Sana3883/compute-optimal_GPT_brain-encoding} .The CNeuroMod dataset is available at \url{https://www.cneuromod.ca/gallery/datasets}


\bibliographystyle{model1-num-names}
\bibliography{sample.bib}
\appendix

\section{Pearson Correlation of Brain Encoding Across All Layers of Block 4 in GPT model for Sub-01}

\label{Appendix_1}
\begin{table}[H]
\centering
\scriptsize
\caption{Layers of block 4 and their corresponding activation shapes}
\renewcommand{\arraystretch}{1.5} 
\begin{tabular}{|c|c|c|}
    \hline 
    Index & Name of layer & Activation shape \\
    \hline \hline
    1 & attn\textunderscore stack.attn\textunderscore nets.4.pre\textunderscore attn\textunderscore norm & [1, 4, 8, 8, 576] \\
    \hline
    2 & attn\textunderscore stack.attn\textunderscore nets.4.post\textunderscore attn\textunderscore dp & [1, 4, 8, 8, 576] \\
    \hline
    3 & attn\textunderscore stack.attn\textunderscore nets.4.attn.w\textunderscore qs & [1, 4, 8, 8, 576] \\
    \hline
    4 & attn\textunderscore stack.attn\textunderscore nets.4.attn.w\textunderscore ks & [1, 4, 8, 8, 576] \\
    \hline
    5 & attn\textunderscore stack.attn\textunderscore nets.4.attn.w\textunderscore vs & [1, 4, 8, 8, 576] \\
    \hline
    6 & attn\textunderscore stack.attn\textunderscore nets.4.attn.fc & [1, 4, 8, 8, 576] \\
    \hline
    7 & attn\textunderscore stack.attn\textunderscore nets.4.attn.attn & [1, 4, 4, 8, 8, 144] \\
    \hline
    8 & attn\textunderscore stack.attn\textunderscore nets.4.pre\textunderscore enc\textunderscore norm & [1, 4, 8, 8, 576] \\
    \hline
    9 & attn\textunderscore stack.attn\textunderscore nets.4.post\textunderscore enc\textunderscore dp & [1, 4, 8, 8, 576] \\
    \hline
    10 & attn\textunderscore stack.attn\textunderscore nets.4.enc\textunderscore attn.w\textunderscore qs & [1, 4, 8, 8, 576] \\
    \hline
    11 & attn\textunderscore stack.attn\textunderscore nets.4.enc\textunderscore attn.w\textunderscore ks & [1, 4, 8, 8, 576] \\
    \hline
    12 & attn\textunderscore stack.attn\textunderscore nets.4.enc\textunderscore attn.w\textunderscore vs & [1, 4, 8, 8, 240] \\
    \hline
    13 & attn\textunderscore stack.attn\textunderscore nets.4.fc & [1, 4, 8, 8, 576] \\
    \hline
    14 & attn\textunderscore stack.attn\textunderscore nets.5.enc\textunderscore attn.attn & [1, 4, 4, 8, 8, 60] \\
    \hline
    15 & attn\textunderscore stack.attn\textunderscore nets.4.pre\textunderscore fc\textunderscore norm & [1, 4, 8, 8, 576] \\
    \hline
    16 & attn\textunderscore stack.attn\textunderscore nets.4.post\textunderscore fc\textunderscore dp & [1, 4, 8, 8, 576] \\
    \hline
    17 & attn\textunderscore stack.attn\textunderscore nets.4.fc\textunderscore block & [1, 4, 8, 8, 576] \\
    \hline
    18 & attn\textunderscore stack.attn\textunderscore nets.4.fc\textunderscore block.0 & [1, 4, 8, 8, 2304] \\
    \hline
    19 & attn\textunderscore stack.attn\textunderscore nets.4.fc\textunderscore block.1 & [1, 4, 8, 8, 2304] \\
    \hline
    20 & attn\textunderscore stack.attn\textunderscore nets.4.fc\textunderscore block.2 & [1, 4, 8, 8, 576] \\
    \hline
\end{tabular}
\label{Table:layers}
\end{table}

\begin{figure}[H]
\centering
\includegraphics[width=0.7\textwidth]{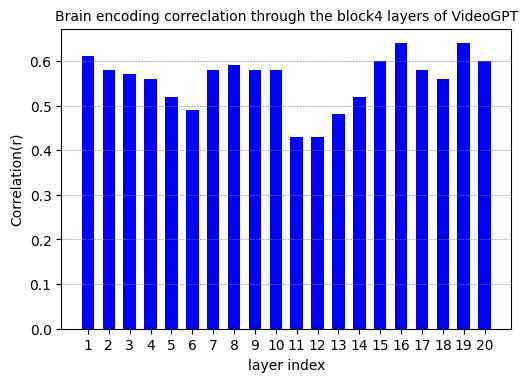} 
\caption{Pearson correlation prediction values for Sub-01 brain encoding across layers presented in Table~\ref{Table:layers}}
\label{Brain_Pearson_Correlation_layers}
\end{figure}






% \end{thebibliography}

\end{document}